\documentclass[conference, hidelinks, table]{IEEEtran}
\IEEEoverridecommandlockouts
\usepackage{cite}
\usepackage{flushend}
\usepackage[all=normal, paragraphs=tight, floats=tight, mathspacing=tight]{savetrees}

\ifCLASSINFOpdf
  \usepackage[pdftex]{graphicx}
  \graphicspath{{figures/}}
  \DeclareGraphicsExtensions{.eps,.jpeg,.png}
\else
  \usepackage[dvips]{graphicx}
  \graphicspath{{figures/}}
  \DeclareGraphicsExtensions{.eps,.jpeg,.png}
\fi

\usepackage{amsmath, amsfonts, bm, amsthm, amssymb}
\usepackage{algorithmic}
\usepackage[ruled,vlined,linesnumbered]{algorithm2e}

\usepackage{array}
\usepackage{blkarray}
\usepackage{tabularx,booktabs}
\newcommand{\ra}[1]{\renewcommand{\arraystretch}{#1}}

\ifCLASSOPTIONcompsoc
 \usepackage[caption=false,font=normalsize,labelfont=sf,textfont=sf]{subfig}
\else
 \usepackage[caption=false,font=footnotesize]{subfig}
\fi

\usepackage{dblfloatfix}

\usepackage{url}

\usepackage{enumitem}

\newlist{inlineroman}{enumerate*}{1}
\setlist[inlineroman]{afterlabel=~,label=\roman*)}
\newcommand{\inlinerom}[1]{
\begin{inlineroman}
#1
\end{inlineroman}
}

\newtheorem{thm}{Theorem}

\newtheorem{prob}{Problem}

\begin{document}
\bstctlcite{IEEEexample:BSTcontrol}
\title{Flex-Net: A Graph Neural Network Approach to\\
Resource Management in Flexible Duplex Networks}


\author{\IEEEauthorblockN{Tharaka Perera\IEEEauthorrefmark{1},
Saman Atapattu\IEEEauthorrefmark{1}, Yuting Fang\IEEEauthorrefmark{1},
Prathapasinghe Dharmawansa\IEEEauthorrefmark{2}, and
Jamie Evans\IEEEauthorrefmark{1}}%
\IEEEauthorblockA{\IEEEauthorrefmark{1}Department of Electrical and Electronic Engineering, University of Melbourne, Australia. \\
\IEEEauthorrefmark{2} Department of Electronic and Telecommunication Engineering,
University of Moratuwa, Sri Lanka.
}
}


%


\maketitle


\begin{abstract}

Flexible duplex networks allow users to dynamically employ uplink and downlink channels without static time scheduling, thereby utilizing the network resources efficiently. This work investigates the sum-rate maximization of flexible duplex networks. In particular, we consider a network with pairwise-fixed communication links. Corresponding combinatorial optimization is a non-deterministic polynomial (NP)-hard without a closed-form solution. In this respect, the existing heuristics entail high computational complexity, raising a scalability issue in large networks. Motivated by the recent success of Graph Neural Networks (GNNs) in solving NP-hard wireless resource management problems, we propose a novel GNN architecture, named \textbf{Flex-Net}, to jointly optimize the communication direction and transmission power. The proposed GNN produces near-optimal performance meanwhile maintaining a low computational complexity compared to the most commonly used techniques.
Furthermore, our numerical results shed light on the advantages of using GNNs in terms of sample complexity, scalability, and generalization capability.

\end{abstract}


%
\IEEEpeerreviewmaketitle

\section{INTRODUCTION}


Recently, time division duplex (TDD) methods have become more popular compared to frequency division duplex (FDD) methods \cite{liu2015performance}, since TDD methods perform better when uplink and downlink data rates are asymmetric. In this paper, we investigate the resource allocation of TDD networks which allow users only to transmit or receive at a given time slot, i.e., half-duplex. Typically in a TDD network, uplink and downlink time slots are predefined. This predefined nature can be relaxed to increase the utilization of the network by allowing dynamic scheduling of communication direction. Such networks are known as flexible duplex networks \cite{dayarathna2020centralized}.

The advantages of dynamic scheduling come at a cost. It introduces a challenging combinatorial optimization problem that is mathematically challenging and computationally expensive to solve. There are various algorithms suggested in the literature for resource scheduling in flexible duplex networks. Previous works include iterative pattern search algorithms for resource scheduling in flexible duplex networks \cite{Dayarathna2021, dayarathna2020centralized}, radio frame selection algorithm for flexible duplex networks \cite{liu2015performance}, a flexible duplex framework for joint uplink and downlink resource allocation \cite{liao2017dynamic}, and resource management with flexible duplex in Narrowband Internet of things (NB-IoT) \cite{malik2019radio}, to name a few. In \cite{popovski2014interference}, authors propose a flexible duplex network with fixed node pairs under the assumption of a balanced traffic load. In this paper, we focus on a similar but more generalized system model than in \cite{popovski2014interference}, which allows the nodes to transmit, receive or be silent depending on the availability of data in the buffer to transfer.

Recent breakthroughs in machine learning on non-Euclidian graph data have also attracted the interest of the wireless network community. The inherent graph structure of wireless networks makes GNNs more suitable than fully-connected neural networks (FCNNs) or convolutional neural networks (CNNs) to tackle wireless network problems. 
Recent efforts using GNNs, such as power control \cite{Shen2019}, channel estimation \cite{tekbiyik2021channel}, cellular network traffic prediction \cite{zhao2020cellular}, and network localization \cite{yan2021graph}, have demonstrated promising results, outperforming classical methods while also offering significant computational complexity improvements.

Motivated by the recent successes of GNNs in wireless networks, for the first time we investigate the potential of using GNNs to jointly optimize the power and communication direction of a flexible duplex network.


The main contributions of this paper are as follows:

\begin{enumerate}[label=\roman*)]
    \item We formulate a novel graph structure that can represent the flexible duplex network. This graph can represent desired links and potential interference links including the direction to efficiently learn the geometric and numerical features of the flexible duplex network. 
    \item We propose a novel GNN model called \textbf{Flex-Net} with an unsupervised-learning strategy to jointly optimize communication direction and transmit power to maximize the sum-rate of the flexible duplex network.
    \item We compare numerical results obtained by extensive simulations using the proposed GNN with baselines listed in Table \ref{table:approaches}. We show that the proposed method outperforms baselines in terms of performance and time complexity. Furthermore, we analyze the sample complexity, scalability, and generalization capability of the proposed approach.
\end{enumerate}
\vspace{-0.5em}
\section{System Model}

\begin{figure}[!tb]
    \centering
    \includegraphics[width=\linewidth]{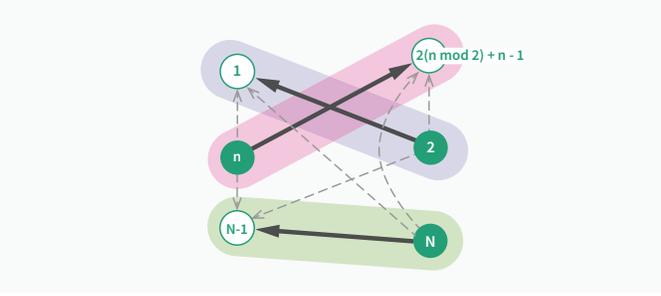}
    \caption{A flexible half-duplex network with $2N$ transceiver nodes.
    }
    \label{fig:network6}
\end{figure}

We consider a flexible duplex network consisting of $2N$ transceiver nodes with interference channels as shown in Fig.~\ref{fig:network6}. In the figure, solid lines and dashed lines indicate desired links and interference links, respectively. Solid circles and hollow circles indicate transmitters (Txs) and receivers (Rxs), respectively.
For convenience, nodes are indexed from $1$ to $2N$. 
Without loss of generality, we define that the nodes with adjacent indices ($1\leftrightarrow2, \dotsc, 2n-1\leftrightarrow{2n}, \dotsc,(2N-1)\leftrightarrow 2N$) act as user pairs. Only one node of each user pair is in Tx mode whereas the other node is in Rx mode.
Overall, there are $N$ user pairs in the network. 
In a Tx-Rx pair, the Rx experiences interference from  ${(N - 1)}$ Tx nodes of other connected pairs.
Most of the existing works consider fixed sets of Tx and Rx \cite{learningToOptimize, Shen2019, liang2019towards}. For convenience, we can consider nodes with odd indices as Rxs and nodes with even indices as Txs. In a network with a fixed set of Tx and Rx pairs, i.e., not a flexible duplex network, the signal to interference-plus-noise ratio (SINR) of a receiver node can be written as
\begin{equation}
    \gamma_{n} = \frac{p_{n+1} |h_{n, n+1}|^2}{\sigma_{n}^2 + \sum_{k \neq n+1}^{N} p_{2k} |h_{n,2k}|^2},
     \; \forall n = 1,3,\dotsc,2N-1,
    \label{eq:sinr_classic}
\end{equation}
where $p_n$ denotes the transmit power of the $n$th node, $h_{n, k}$ denotes the complex channel state information (CSI) from the $k$th Tx to the $n$th Rx and $\sigma_n^2$ denotes the noise power at the $n$th Rx.

In this work, we consider a flexible duplex network where we do not have a fixed set of Tx or Rx in a given user pair. Hence, we need to define the SINR for all the nodes in the network. SINR of the $n$th node of a flexible duplex network is given by
\begin{equation}
    \gamma_n = \frac{p_m d_m |h_{n, m}|^2}{\sigma_n^2 + \sum_{k \neq m, n}^{2N} p_k d_k |h_{n,k}|^2}, \; \forall n = 1,2,\dotsc,2N,
    \label{eq:sinr}
\end{equation}
where  $m = 2(n\bmod2) + n - 1$, $d_m = 1 - d_n$, and $d_n \in \{0, 1\}$. The Rx state and Tx state of the $n$th node are represented by $d_n = 0$ and $d_n = 1$, respectively.  
If a node has no data to transmit, its CSI can be set to zero to ensure that no resources are allocated to it.
Furthermore, we consider a more realistic non-reciprocal channel environment, i.e., $h_{m,n}$ may not be equal to $h_{n,m}$. 


\section{Optimization Problem}
In this section, we formulate the optimization objective of the network, which aims to jointly optimize the power allocation and the communication direction of the flexible duplex network. 
To this end, the sum-rate utility maximization problem can be written as
\begin{prob}\label{eq:optimization_objective}
\begin{equation*}
\begin{split}
\max _{p_{n}, d_{n}, \! \forall n} \quad &\sum_{n=1}^{2N} \log _{2}\left(1+\frac{p_{m} d_m \left|h_{n, m}\right|^{2}} {\sigma_n^{2}+\sum_{k \neq m,n}^{2N} p_{k}  d_{k}\left|h_{n, k}\right|^{2}}\right),
\end{split}
\end{equation*}
\begin{equation*}
\begin{split}
\text { s.t. } \quad &0 \leq p_{n} \leq P_{\max}, \quad \forall n, \\
 &m = 2(n\bmod2) + n - 1, \quad \forall n, \\
 &d_m = 1 - d_n, \quad \forall n, \\
 &d_n \in \{0, 1\}, \quad \forall n, \\
\end{split}
\end{equation*}
\end{prob}
\noindent where $P_{\max}$ denotes the maximum transmit power of each node.
\begin{thm}\label{thm:np_hard}
Problem \ref{eq:optimization_objective} is NP-hard for any number of user pairs.
\end{thm}

\begin{IEEEproof} 
For a network with a fixed set of Tx and Rx pairs, i.e., that is not a flexible duplex network, following the convention that nodes with odd indices are Rxs and nodes with even indices are Txs, the sum-rate maximization problem can be written as \cite{learningToOptimize}
\begin{prob}\label{eq:traditional_optimization_problem}
\begin{equation*}
\begin{split}
    \max _{p_{n}, \; \forall n} \quad &\sum_{n=1}^{N} \log _{2}\left(1+\frac{p_{2n} \left|h_{2n-1, 2n}\right|^{2}} {\sigma_{2n-1}^{2}+\sum_{k \neq n}^{N} p_{2k} \left|h_{2n-1, 2k}\right|^{2}}\right) ,\\
\text { s.t. } \quad &0 \leq p_{n} \leq P_{\max}, \quad \forall n
 \end{split}
\end{equation*}
\end{prob}

\noindent which is known as NP-hard \cite{luo2008dynamic}. There are only $N$ Txs hence the upper bound of the summation. If there exists a polynomial-time reduction from Problem \ref{eq:optimization_objective} to Problem \ref{eq:traditional_optimization_problem}, we can conclude that Problem \ref{eq:optimization_objective} is NP-hard.

In a given flexible duplex network, assume that nodes with even and odd indices act as Tx (${d_n=1}$) and Rx (${d_n=0}$), respectively. Then only the nodes with odd indices will contribute to the sum-rate. Hence, Problem \ref{eq:optimization_objective} can be reduced to Problem \ref{eq:traditional_optimization_problem} by substituting communication directions. There exists a linear-time transformation from Problem \ref{eq:optimization_objective} to Problem \ref{eq:traditional_optimization_problem}. This completes the proof.
\end{IEEEproof}
As mentioned in the Theorem \ref{thm:np_hard}, Problem \ref{eq:optimization_objective} is an NP-hard, non-convex combinatorial optimization problem with binary constraints. We provide two potential ways to obtain near-optimal solutions in the next section.

\section{Joint~power~and communication direction optimization}

In this section, we focus on two methods to solve the optimization Problem \ref{eq:optimization_objective}. There are two variables to optimize, namely  \inlinerom{\item the power allocation vector denoted by $\bm p = [p_1,\dotsc,p_{2N}]^\mathsf{T}$, and \item the communication direction vector denoted by $\bm d = [d_1,\dotsc,d_{2N}]^\mathsf{T}.$}
First, we develop a heuristic approach based on an iterative coordinate descent method.
Then, we propose a novel GNN approach, called \textbf{Flex-Net}, to jointly optimize power allocation and communication directions.

\subsection{Heuristic Approach}

In this approach, we use an iterative coordinate descent method to optimize one variable at a time. Since this is a combinatorial problem with integer variables, we adapt the direct search mechanism suggested in \cite{dayarathna2020centralized} to optimize the discrete integer variable $\bm d$ while keeping $\bm p$ fixed.

First, the communication direction vector $\bm d$ is optimized using the direct search algorithm with initial power values. 
Then the power vector $\bm p$ is recalculated using the weighted minimum mean square error (WMMSE) \cite{shi2011iteratively} algorithm. 
The above steps are repeated until the achieved sum-rate is not improved by a user-defined constant $\varepsilon$ over subsequent iterations. 
Despite the near-optimal sum-rate performance, the computational complexity of this method is $\mathcal{O}(n^4)$\footnote{
The number of initializations increases linearly and the worst-case time complexity of each run is $\mathcal{O}(n^3)$. This includes the linear complexity of sum-rate calculation. Altogether, the time complexity is $\mathcal{O}(n^4)$.
}. This makes the algorithm difficult to be applied for real-time resource allocation in large-scale networks. To overcome the computational complexity issue of this algorithm, we propose a novel GNN-based approach in the next subsection.

\subsection{\textbf{Flex-Net} Approach}
In this subsection, we propose a novel GNN named \textbf{Flex-Net} to jointly optimize $\bm p$ and $\bm d$. 
First, we represent the flexible duplex network as a graph with two types of edges. CSI of desired links and crosslinks are used as vertex features and edge features, respectively. The created graph is used as the input to the proposed GNN model. 
A short introduction about GNNs and details of the graph formulation, proposed architecture, theoretical motivation, and optimization methods are detailed below.


\subsubsection{Graph Neural Networks}

There are two main types of GNNs: Spectral GNNs and Spatial GNNs \cite{bronstein2017geometric, zhang2020deep}. Spectral GNNs are designed based on a signal processing perspective and operate in the spectral domain, while spatial GNNs focus on the structure of the graph and can aggregate information from neighboring nodes like the convolution kernels present in Convolutional Neural Networks (CNNs). 
Being restricted to certain operations in the frequency domain, Spectral GNNs have limited expressive power.
Therefore, we use a spatial GNN to optimize the objective function in Problem \ref{eq:optimization_objective}.

Given a graph structure with node features and edge features, GNNs use neighborhood aggregation functions and combination functions to learn representations of nodes of the graph. This process is followed for multiple iterations to learn sufficient structural information about the neighborhood of nodes. 
Mathematically, aggregation and combination steps of the $\ell$th layer are
\begin{equation*}
    \begin{split}
        &
        \resizebox{\linewidth}{!}{$%
        \bm a_{v}^{(\ell)}=\operatorname{AGGREGATE}^{(\ell)}\hspace{-0.1em}\left(\left\{\bm x_{u}^{(\ell-1)}, \bm x_{v}^{(\ell-1)}, \bm e_{uv}: u \in \mathcal{N}(v)\right\}\right)
        $%
        }%
        \\
        &\bm x_{v}^{(\ell)}=\operatorname{COMBINE}^{(\ell)}\left(\bm x_{v}^{(\ell-1)}, \bm a_{v}^{(\ell)}\right),
    \end{split}
\end{equation*}
where $\bm x_{v}^{(\ell)}$ is the embedding of node $v$ at the $\ell$th layer, $\bm e_{uv}$ is the edge feature between node $u$ and $v$, and $\mathcal{N}(v)$ represents the neighbours of node $v$.

\subsubsection{Graph Representation of the Flexible Duplex Network}

In our system model, there are $2N$ transceiver nodes. The signal from each Tx acts as interference to the unintended Rxs. Observing this physical network structure, we formulate the network graph as follows.

\begin{figure}[tbp]
    \centering
    \includegraphics[width=\linewidth]{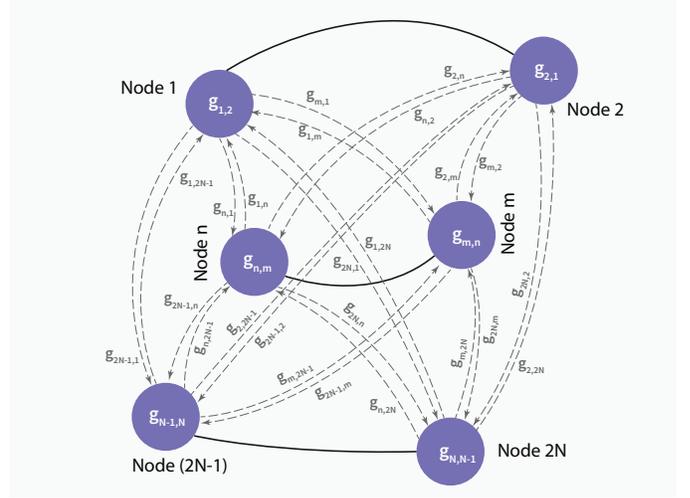}
    \caption{Graph representation of a flexible duplex network. Solid lines indicate $\bm e_d$ and dashed lines indicate $\bm e_i$,
    where $g_{i,j} = |h_{i,j}|^2$ and ${m = 2(n\bmod2) + n - 1}$.}
    \label{fig:graph}
\end{figure}

Each node in the network is represented by a vertex in the graph. Hence, there are $2N$ vertices in the graph. The $n$th vertex contains $g_{n,m} = |h_{n,m}|^2$ as the vertex feature. This is motivated by the fact that the sum-rate expression contains the squared magnitude of CSI in the numerator.
Vertices of the graph are connected using two types of edges. 
\inlinerom{\item Undirected edges represent desired links of the network, denoted by $\bm e_d$, and \item directed edges represent all the \textit{potential} interference, denoted by $\bm e_i$}.
The squared magnitude of CSI from node $v$ to $u$ is used as the edge feature of $\bm e_i$.  This is motivated by the fact that the interference is a function of squared channel magnitudes. 
These $\bm e_i$ edges become an interference only if the origin node of an edge is a Tx, hence the name potential interference. 
Fig.~\ref{fig:graph} illustrates the network graph of a flexible duplex network with $2N$ transceivers. 

\subsection{Motivation for the Proposed Architecture}
There has been growing interest in the literature to explore the architecture of a GNN to maximize its representation capability. Strictly speaking, GNN should construct unique node embeddings which represent the neighborhood and node features. As pointed out in \cite{xu2018powerful} when the node and edge features are from a countable multiset, the aggregation and combination functions should be injective for the GNN to maximize its representation capability. 

From a wireless communication standpoint, there might be multiple nodes that share the same optimized power and direction values despite the differences in associated CSI. This suggests that multiple nodes with different neighborhoods may share the same embedding, i.e., GNN layer $\mathcal{F}: \bm x^{(\ell-1)} \rightarrow \bm x^{(\ell)}$ may take a non-injective form.
In the next subsection, we propose a more flexible GNN architecture, which can approximate any injective or non-injective function. In particular, we relax the constraint of using injective functions. Empirical results suggest that the proposed architecture generalizes well for the objective proposed in Problem \ref{eq:optimization_objective}.

\subsection{\textbf{Flex-Net} Architecture}
In general, GNNs can perform three types of tasks
\inlinerom{\item node-level, \item edge-level, and \item graph-level.}
In node-level tasks, predictions are based on individual nodes. As the name suggests edge-level tasks are performed to find the presence, direction, or any other properties of the edges. Graph-level tasks are performed on the entire graph to obtain insights into the whole graph.

In this work, we perform two types of tasks on the network graph. An edge-level task is used to find the directions of the desired links between nodes. In addition to that, a node-level task is used to predict the optimal power value for each node. Our GNN is comprised of two information aggregation steps to exploit connections represented by 
$\bm e_d$ and $\bm e_i$. We perform the following operations on two types of edges:
\begin{equation}
\begin{alignedat}{3}
    &\bm \alpha^{(\ell)}_{v;\text{intf}} &&= 
    \gamma \, &&\left[ \phi \left( \bm W^{(\ell)}_{u;\text{intf}} \bm{x}_u^{(\ell-1)} +\bm W^{(\ell)}_{v;\text{intf}} \bm{x}_v^{(\ell-1)} \right. \right.  \\
    &&&&&\qquad \qquad \qquad \qquad \left. \left. +\bm W^{(\ell)}_{e;\text{intf}} \bm{e}_{u,v} : u \in \mathcal{N}_\text{intf}(v) \right) \right],\\
    &\bm c_v^{(\ell)} &&= 
       &&\left( \bm x_v^{(1)} \parallel \bm \alpha^{(\ell)}_{v;\text{intf}} \right), \\
    &\bm \alpha^{(\ell)}_{v;\text{dsr}} &&= \gamma \, &&\left[ \phi \left( \bm W^{(\ell)}_{u;\text{dsr}} \bm{c}_u^{(\ell)}+\bm W^{(\ell)}_{v;\text{dsr}} \bm{c}_v^{(\ell)} : u \in \mathcal{N}_{dsr}(v) \right) \right],\\
    &\bm x_v^{(\ell)} &&= 
       &&\left( \bm x_v^{(1)} \parallel \bm \alpha^{(\ell)}_{v;\text{dsr}} \right),
\end{alignedat}
\end{equation}
where $\gamma$ represents a permutation invariant pooling function such as maximum or summation, $\phi$ represents a non-linear activation function, $\bm{x}_v^{(\ell)}$ represents the node features of the node $v$ in the $\ell$th layer, $\bm W^{(\cdot)}_{(\cdot;\cdot)}$ matrices represent trainable weight matrices, $\mathcal{N}_\text{intf}(v)$ is the set of adjacent nodes connected with $\bm e_i$ edges towards node $v$, $\mathcal{N}_\text{dsr}(v)$ is the set of adjacent nodes connected to node $v$ with $\bm e_d$ edges, $\bm{e}_{u, v}$ denotes edge feature of $\bm e_i$ from node $u$ to $v$ and $\bm x^{(1)}_v$ denotes the node embedding of the first layer which is the squared channel magnitude of the desired link. Skip connection from the first layer to layer $\ell$ is called a residual connection. Such residual connections reduce the risk of vanishing or exploding gradients during backpropagation. Empirically, residual connections result in a significant increase in the performance of deep neural networks \cite{he2016deep}.
The symbol $\parallel$ denotes the vector concatenation. 
Moreover, unlike the GNNs described in previous research, the \textbf{Flex-Net} is able to accurately generate node embeddings for a flexible duplex network due to its use of two aggregation steps.

After performing the above aggregation and combination steps for multiple layers, embeddings of the final layer are used to predict the power values of nodes as follows:
\begin{equation}
    p_n = P_{\max} \operatorname{SIGMOID} \left( \frac{1}{T_p}\operatorname{MLP} \left( \bm{x}_n^{(\text{final})} \right) \right),
\end{equation}
where $T_p$ is a scaling parameter called the \textit{temperature}, $\operatorname{MLP}$ represents a trainable multi-layer perceptron, and $\bm x_n^{(\text{final})}$ represents the final layer embedding of $n$th node.  Usage of a low-temperature value provides more extreme power values biased towards $0$ or $P_{\max}$ which are similar to the power values obtained using the WMMSE algorithm.


Directions of the desired links are found by considering the embeddings of adjacent nodes connected by $\bm e_d$. As described in the system model, the direction is represented with a binary variable. Due to the lack of differentiability, it is difficult to optimize the network with the binary constraint. Therefore, we relax the binary constraint and consider it to be a real number between $0$ and $1$. Then we decide the direction of the edge by considering the embeddings of nodes in both ends as follows:
\begin{equation}
    d_{u,v} =  \operatorname{SIGMOID} \left( \frac{1}{T_d} \operatorname{MLP}  \left( \bm{x}^{(\text{final})}_{u} \parallel \bm{x}^{(\text{final})}_{v} \right) \right),
\end{equation}
where $T_d$ is the temperature parameter, $\bm{x}_u^{(\text{final})}$ and $\bm{x}_v^{(\text{final})}$ denotes node embeddings of the adjacent nodes connected by $\bm e_d$. Similar to the case of power allocation, usage of a low-temperature value works as a regularizer to restrain the relaxed binary variable to a tight neighborhood of $0$ or $1$.

We optimize the trainable parameters of the network using \textit{ADAM} which is an adaptive variant of the stochastic gradient descent algorithm. We use the negative value of Problem~\ref{eq:optimization_objective} with relaxed binary constraint as the loss function. This unsupervised learning mechanism eliminates the need of labels for training. Data (channel realizations) required for the training process is generated by following the method explained in Section \ref{sec:numerical_results}.




\section{Numerical Results}\label{sec:numerical_results}

In this section, we compare the proposed approach with different existing algorithms in terms of performance, time complexity, and generalization capability. 


\begin{figure*}[!t]
\centering
\subfloat[]{\includegraphics[width=0.26\linewidth]{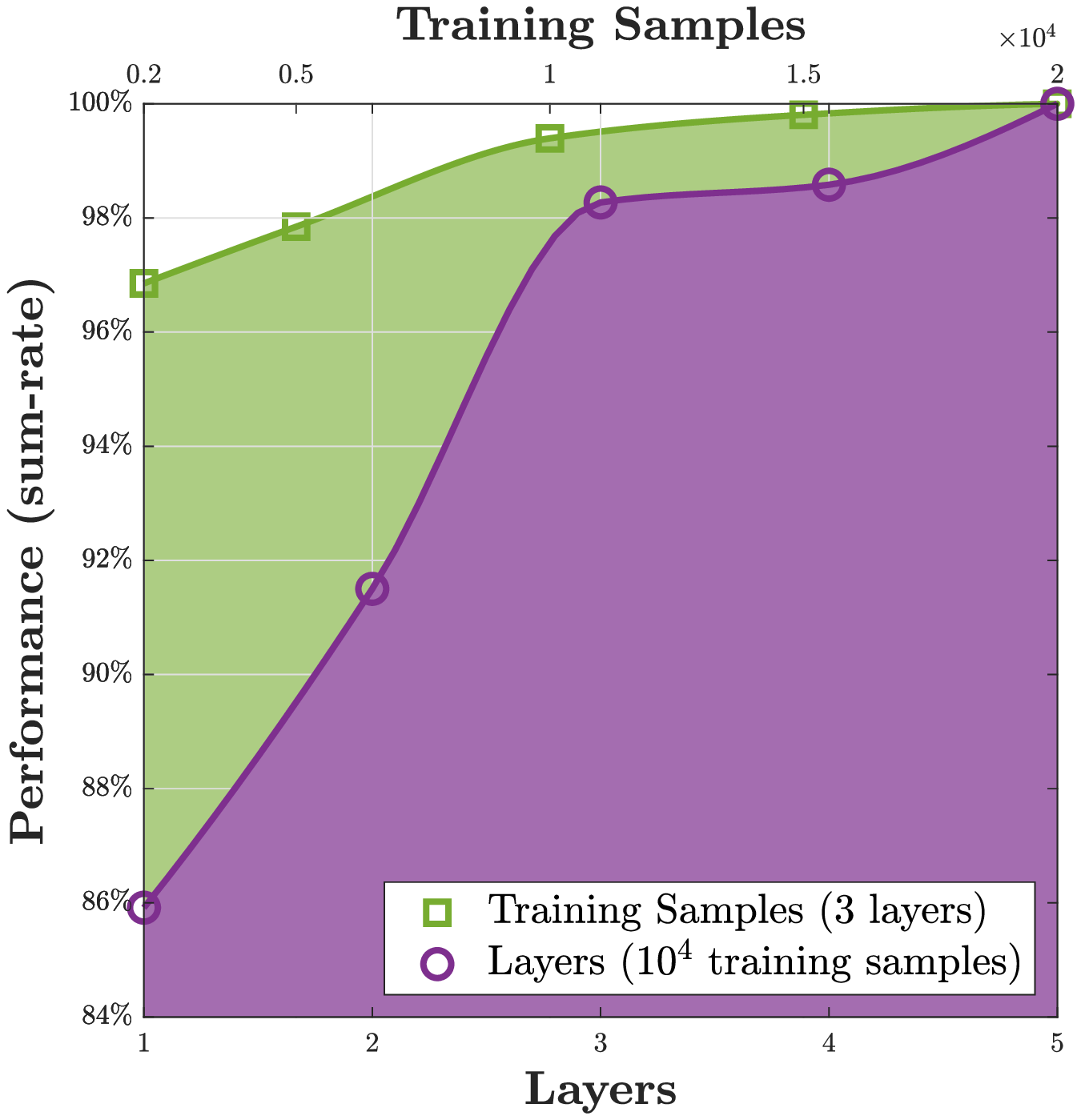}
\label{fig:samples_and_layers}}
\hfil
\subfloat[]{\includegraphics[width=0.265\linewidth]{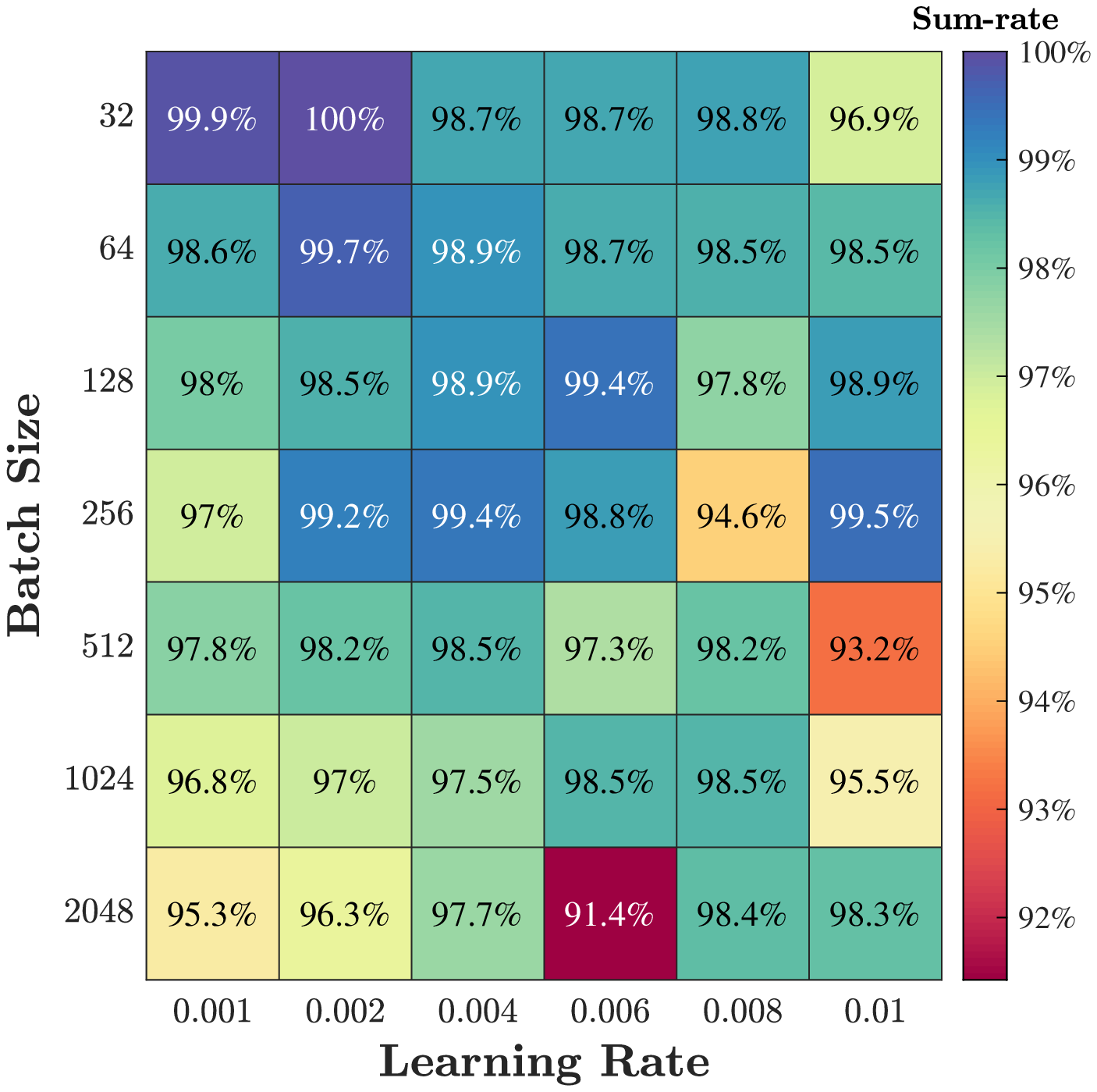}
\label{fig:batch_and_lr}}
\hfil
\subfloat[]{\includegraphics[width=0.263\linewidth]{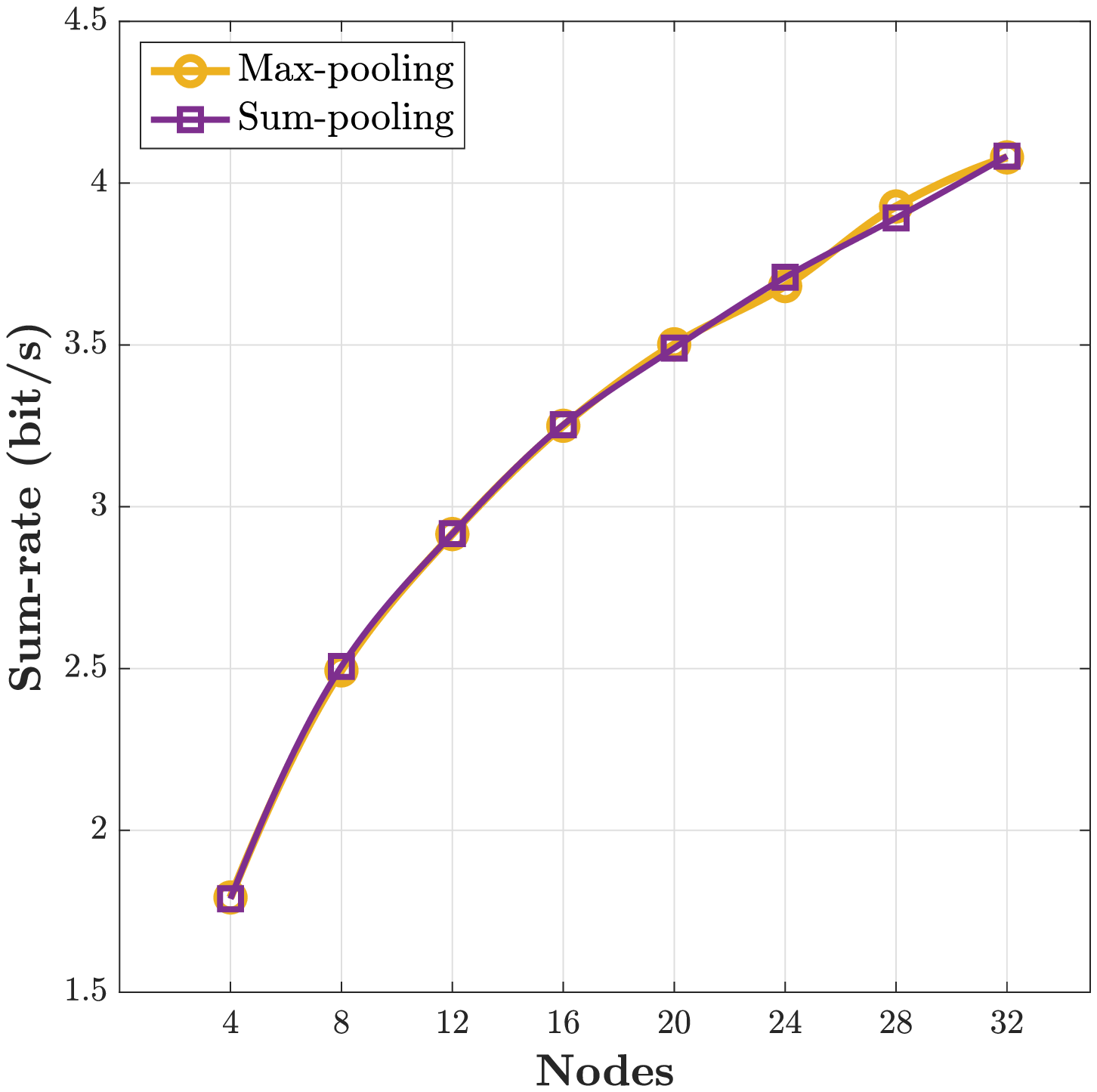}
\label{fig:pooling}}
\caption{Hyperparameter tuning of the \textbf{Flex-Net}. (a) Average sum-rate performance against the sample count and the number of layers. (b) Average sum-rate performance against mini-batch size and learning rate. (c) Average sum-rate comparison of max-pooling and sum-pooling. }
\end{figure*}

\begin{figure*}[!t]
\centering
\subfloat[]{\includegraphics[width=0.26\linewidth]{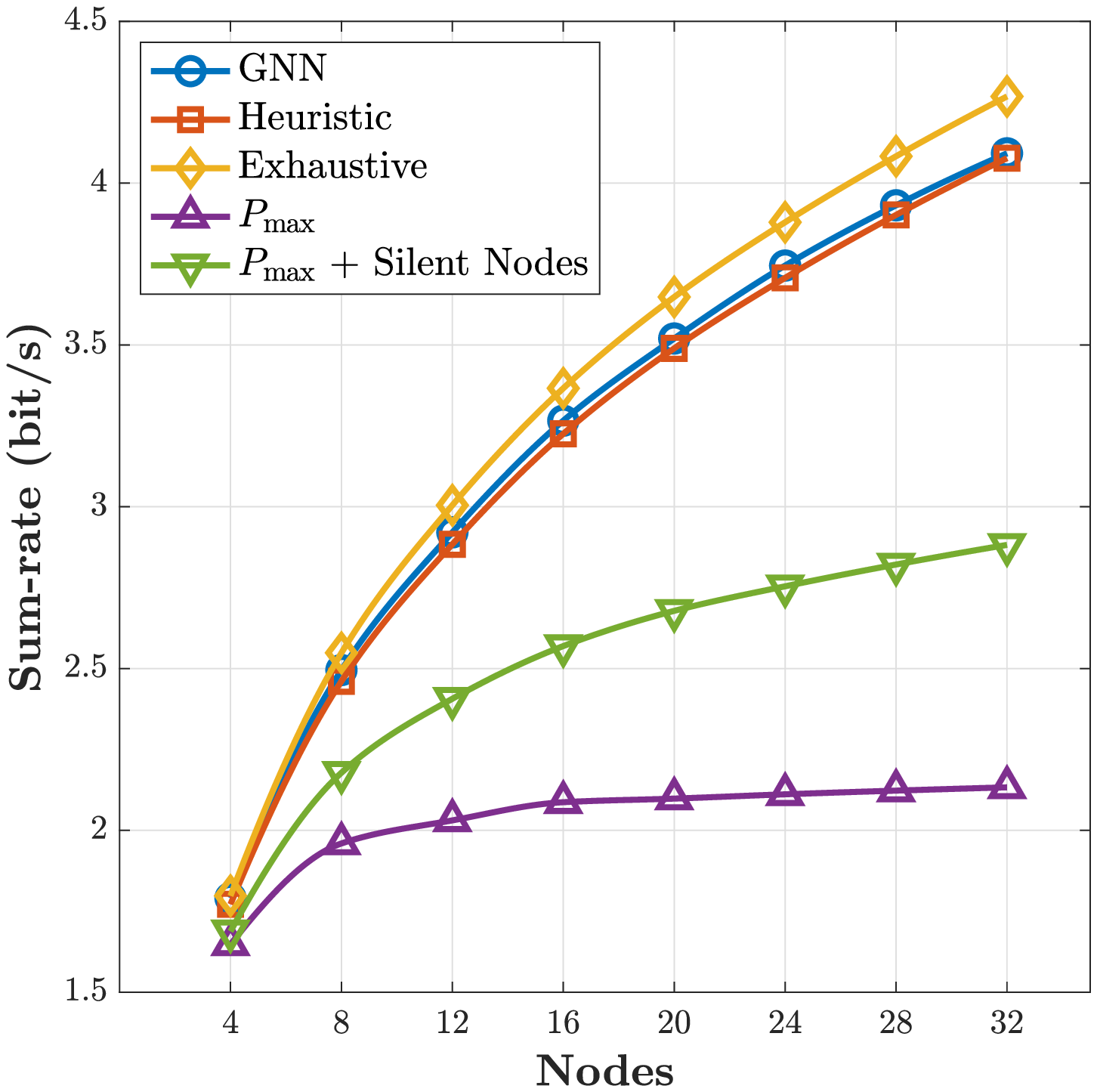}
\label{fig:performance}}
\hfil
\subfloat[]{\includegraphics[width=0.265\linewidth]{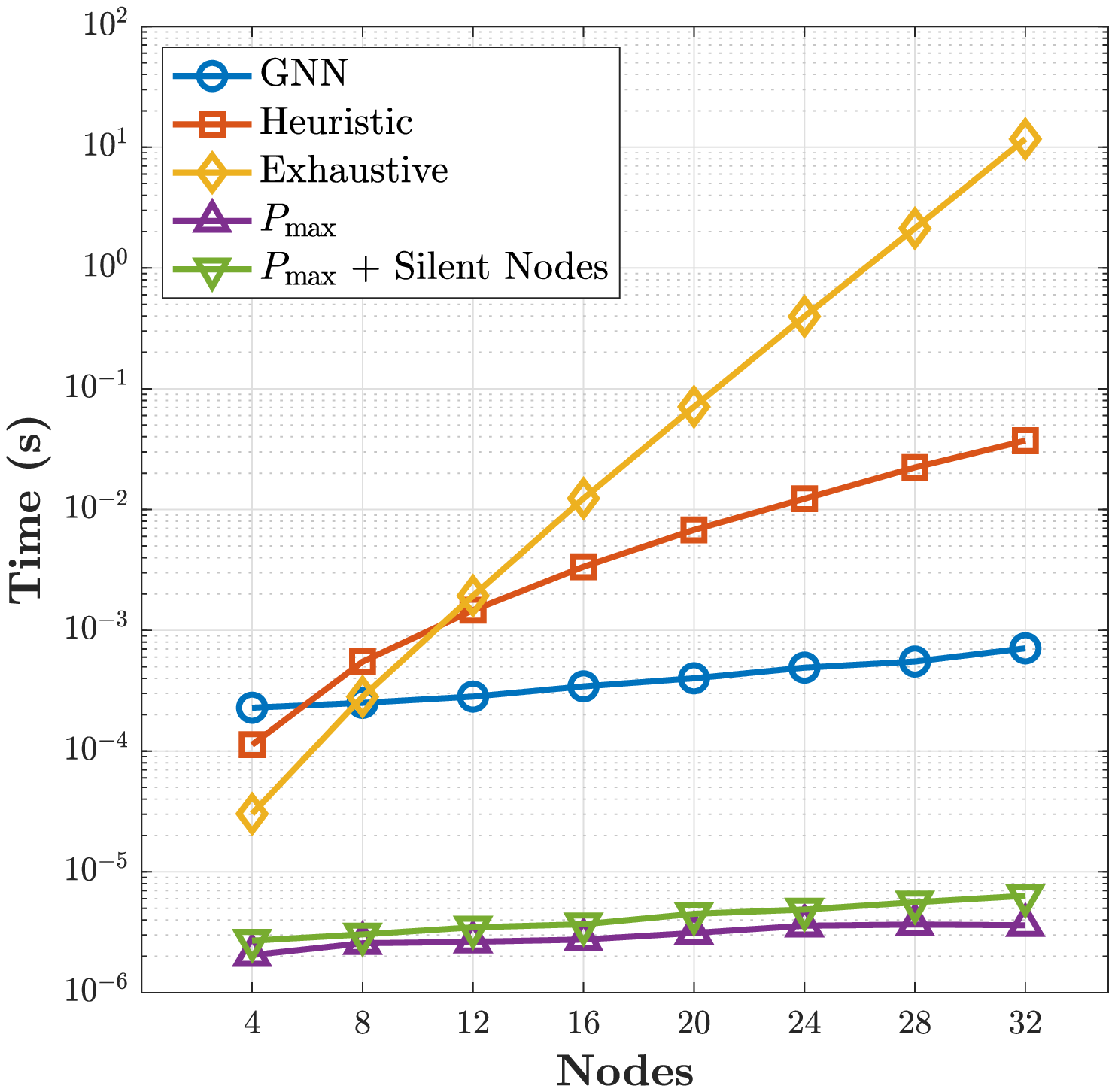}
\label{fig:time_complexity}}
\hfil
\subfloat[]{\includegraphics[width=0.26\linewidth]{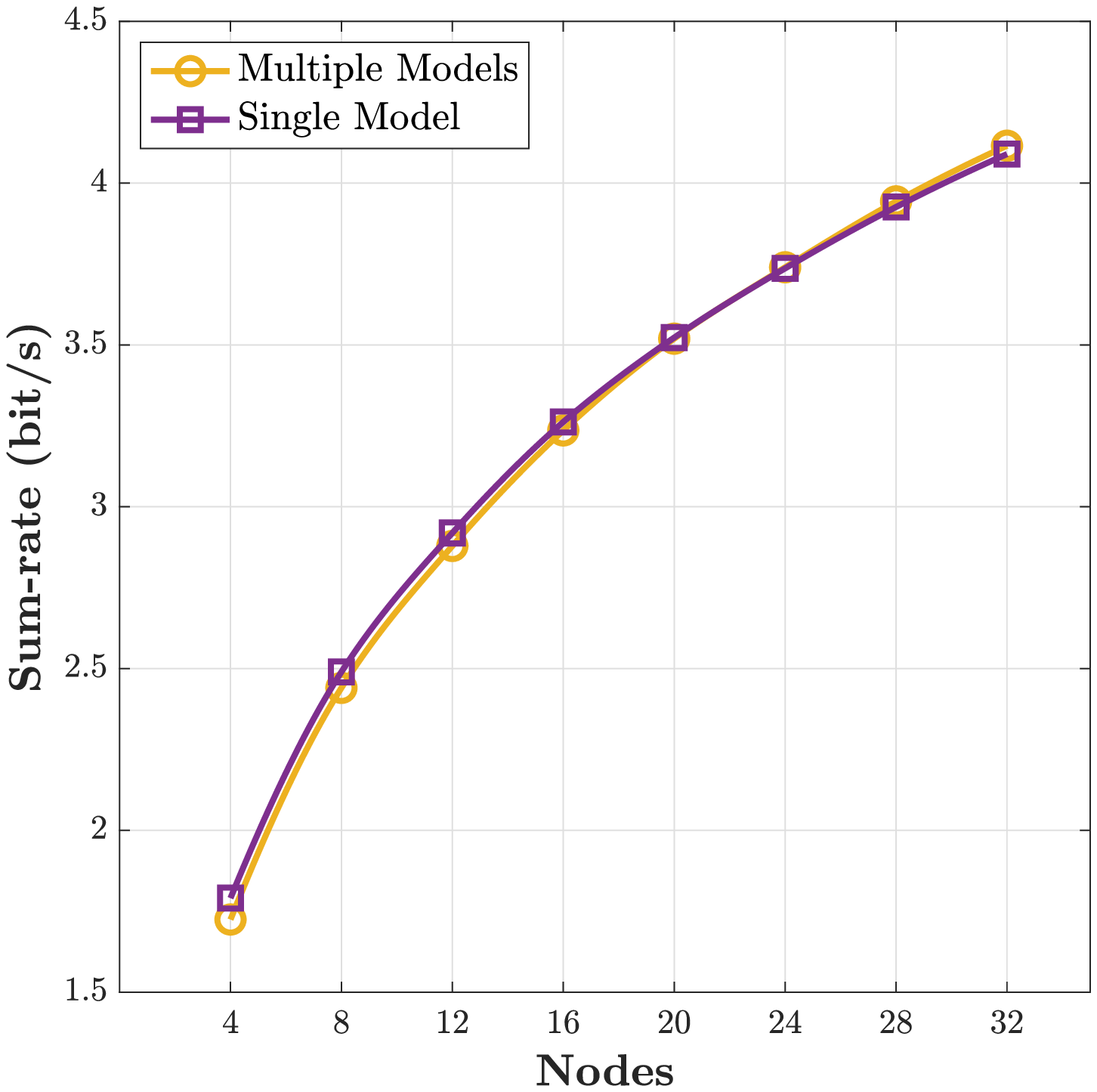}
\label{fig:extrapolation}}
\caption{(a) Average sum-rate comparison of different algorithms. (b) Running time per sample for different algorithms. (c) Sum-rate comparison of a set of models (multiple models) which are trained with a fixed number of nodes and a model (single model) trained with a variable number of nodes. }
\end{figure*}

\begin{table}[!b]
\centering
\caption{Algorithms used in the experiments}
\ra{1.3}
\begin{tabular}{@{}llr@{}}\toprule
Approach &  Time Complexity &  Performance (avg.) \\ \midrule
Exhaustive Search & $\mathcal{O}(2^n)$ & 100\% \\
\textbf{Flex-Net} & $\bm{\mathcal{O}(n^2)}$ & \textbf{95.8\%} \\
Heuristic Search & $\mathcal{O}(n^4)$ & 95.5\% \\
Max Power & $\mathcal{O}(n)$ & 49.9\% \\
Max Power with Silent Nodes & $\mathcal{O}(n^2)$ & 67.5\%

\\ \bottomrule
\end{tabular}
\label{table:approaches}
\end{table}
\subsection{Simulation Setup}
We consider a flexible duplex network spanned over ${4 \times 4 \; \text{km}^2}$ area.
For convenience, we assume the devices are arranged in a 2D space, but the devices can be distributed in any dimensional space in our GNN model.
We assume the devices are Poisson disk distributed which is similar to a uniformly distributed case but with a minimal distance of $100\,$m. Devices are paired randomly to form desired links. 

To generate CSI, we consider the large and small-scale fading effects of the network \cite{goldsmith2005wireless}. Free-space path loss is calculated for $5\,$GHz frequency with log-normal shadowing of $9.5\,$dB. Furthermore, we assume a Rayleigh fading channel which is a suitable model to simulate fading in an urban environment and is commonly used for performance evaluation of resource allocation methods.

Simulations are carried out for 4-user to 32-user scenarios. We compare the proposed approach with the baselines given in table \ref{table:approaches}. 
For the \textit{Exhaustive Search}, the WMMSE algorithm is applied to all the direction combinations.
The \textit{Max power} strategy selects the direction with the strongest CSI magnitude as the communication direction and uses $P_{\max}$ as the transmit power. 
In addition to that, \textit{Max power with Silent Nodes} strategy turns off the transmitter if there is at least one crosslink with twice the received power of the desired link.
All the algorithms are implemented in Python language
\footnote{The code is available at \url{https://github.com/tharaka-perera/flex-net}}.
For large matrix manipulations, Numpy is used. 

All the implemented classical algorithms are accelerated to achieve native machine code performance using a just-in-time compiler.
Neural networks are implemented from the scratch using Pytorch and Pytorch Geometric frameworks. To do a fair comparison, all the processing is done on a single-core CPU during performance and time comparisons. GPU is used only during training to reduce training time.
\subsection{Hyperparameter Tuning of the GNN}
Neural networks comprise different hyperparameters. In our experiments, we analyze the effect of a few key hyperparameters. 

In the literature \cite{xu2020can}, it is considered that GNNs demand a smaller number of training samples compared to conventional FCNNs. To check the validity of that hypothesis and to find a suitable training sample count for our experiments, we test the performance against the number of training samples for the 32-user case. Each model is trained for 50 iterations and the performance is evaluated. Results are shown in Fig. \ref{fig:samples_and_layers}. The proposed model achieves more than 99\% of performance with $10^4$ samples. This is significantly low compared to the sample complexity of power control approaches proposed with FCNNs or CNNs \cite{liang2019towards, zhao2020power}.

Furthermore, we evaluate the effect of the number of layers in the GNN for performance. Results are presented in Fig. \ref{fig:samples_and_layers}. Even though the performance is improved with the number of layers, \textbf{Flex-Net} can outperform the classical approach with only 3 GNN layers. We use 3 layers in the rest of the simulations.

Finally, we select the mini-batch size and learning rate using a parametric grid-search. Results are illustrated in Fig.~\ref{fig:batch_and_lr}.
Small mini-batch sizes combined with small learning rates result in better performance. In contrast, higher learning rates results in overshooting, hence the instability and poor performance. We choose the mini-batch size of 64 (improved training time) with 0.002 learning rate for the rest of the simulations.

\subsection{Pooling Function}

In the proposed GNN architecture, node embeddings are updated recursively layer by layer. As the pooling function of the aggregation function, the usage of an injective multiset function, \textit{sum-pooling} is preferred over \textit{max-pooling} \cite{xu2018powerful}; nevertheless, succeeding work such as \cite{garg2020generalization} questions this argument. In our experiments, we find that the performance of both pooling functions is almost identical. We use \textit{sum-pooling} in rest of the experiments.


\subsection{Performance}

The average sum-rate performance of the proposed architecture is compared against all the baseline approaches. The simulation results are summarized in Fig. \ref{fig:performance}. Simulation results indicate that the proposed GNN approach outperforms all the classical methods in the network configurations tested.
Moreover, it can achieve tightly close results to that of the exhaustive search. 


\subsection{Time Complexity}

 In the midst of growing demand for higher data rates with lower latency, it is crucial for the wireless network to control thousands of users with minimal processing overhead. GNN shines in this regime with its capabilities to match the performance of classical algorithms while keeping the time complexity significantly low. We compare the average running time of GNN with other baselines for $10^4$ test samples. Simulation results are given in Fig. \ref{fig:time_complexity}. We can see that GNN outperforms both the exhaustive search and the heuristic search significantly. This makes the GNN approach suitable to be used in real-time networks with low latency.


\subsection{Generalization}

GNN operations neither depend on the number of nodes nor the number of edges in the network. Hence GNNs can handle variable input sizes without retraining, unlike traditional FCNNs. This makes it incredibly practical to be used in wireless networks with a variable number of users. 

To study the generalization capability of GNN, instead of training multiple models for different numbers of nodes, we train a single model and test performance for networks with 4 to 32 users. Samples with different numbers of nodes are used for the training. Empirical data presented in Fig. \ref{fig:extrapolation} suggests the performance of \textbf{Flex-Net} is not affected by the changes in the number of users.

\section{Conclusion}

In this work, a joint power allocation and communication direction selection problem is investigated for flexible duplex networks. 
We considered a system model that can dynamically schedule the communication direction of a TDD network. With the goal of maximizing the sum-rate performance of the network, an optimization problem is formulated. We prove that the optimization problem in focus is an NP-hard problem. To circumvent this challenging problem, a novel GNN-based approach is presented. First, a flexible duplex network is represented as a graph to be used as the input to the proposed GNN model. Then, the GNN model is optimized using the generated graph data in an unsupervised manner. To obtain the maximum network utility possible, a grid-search method is used to find suitable hyperparameters for the GNN. Finally, extensive numerical analysis is carried out to compare the performance of the proposed approach and four other baselines. Numerical results suggest that the proposed \textbf{Flex-Net} method can generate near-optimal results with reduced time complexity compared to existing methods. Further analysis verified the advantages of the proposed method in terms of sample complexity, scalability, and generalization capability.


\bibliographystyle{IEEEtran}
\bibliography{IEEEabrv,refs}

\end{document}